# Twenty Hirsch index variants and other indicators giving more or less preference to highly cited papers

Michael Schreiber

*Institut für Physik, Technische Universität Chemnitz, 09107 Chemnitz, Germany*
*E-mail: schreiber@physik.tu-chemnitz.de*
*Phone: +49 371 531 21910, Fax: +49 371 531 21919*

**Abstract.** – The Hirsch index or $h$-index is widely used to quantify the impact of an individual's scientific research output, determining the highest number $h$ of a scientist's papers that received at least $h$ citations. Several variants of the index have been proposed in order to give more or less preference to highly cited papers. I analyse the citation records of 26 physicists discussing various suggestions, in particular $A$, $e$, $f$, $g$, $h(2)$, $h_w$, $h_T$, $\hbar$, $m$, $\pi$, $R$, $s$, $t$, $w$, and maxprod. The total number of all and of all cited publications as well as the highest and the average number of citations are also compared. Advantages and disadvantages of these indices and indicators are discussed. Correlation coefficients are determined quantifying which indices and indicators yield similar and which yield more deviating rankings of the 26 datasets. For 6 datasets the determination of the indices and indicators is visualized.

## 1 Introduction

In spite of fundamental reservations bibliometrics methods are used more and more often for evaluation purposes. Not only administrators but also selection committees use scientometric data in their assessment, since large electronic data bases enable a reasonably fast determination of publication lists and corresponding citation records. Thus very crude indicators like the number of publications of a scientist can be replaced by more comprehensive information like the citation frequency of each publication. But for a comparison of different datasets the dangerous idea to quantify the research output by a single number remains fascinating. Simple indicators as the total number of citations to all papers or the average citation frequency have obvious disadvantages like the difficulty to determine all the citation counts with reasonable accuracy or giving undue weight to highly cited review articles, or taking a possibly large number of irrelevant (not or lowly cited) papers into account. This can be avoided by considering only a small number of relevant or significant papers, but this solution raises the question how to determine this core set of significant papers from a given set of publications. Taking a fixed number or a certain percentage of all publications into consideration would mean a somewhat arbitrary and biased choice.

An elegant solution of this difficulty was proposed by Hirsch [1] designing the now famous or – depending on the point of view – infamous $h$ index. It is defined as the highest number of papers of a scientist that have been cited $h$ or more times. In this way the size $h$ of the core is determined self-consistently. Moreover, it is relatively easy to determine $h$ with reasonable accuracy, because only a rather small number of publications has to be taken into account. Nevertheless there have been suggestions to reduce the number of significant publications even further, e.g. by requiring that the citation frequency in the core should be larger than or equal to the squared number [2] or ten times the number [3] of papers in the core, thus giving more preference to highly cited papers.

These variants do not solve a major disadvantage of the $h$ index, namely that it does not take into account the skewness of the citation distributions. Already in his original publication [1] Hirsch noted that "there is considerable variation in the skewness" so that "for an author with ... a few seminal papers with extraordinarily high citation counts, the $h$ index will not fully reflect that scientist's accomplishments". The problem can also be formulated in the following way: Once a paper has reached the number of citations which qualifies it for the core set, then further citations are irrelevant. It is straightforward to treat this difficulty by considering the average number of citations to the papers in the core. This could be done for the $h$ core leading to the complementary index $A$ [4] or for the elite set suggested by Vinkler [5]. The $g$ index which was proposed by Egghe [6] captures the spirit of the $h$ index in so far as its core size can also be self-consistently defined as the highest number $g$ of papers of a scientist that have been cited $g$ or more times on average [7], where the arithmetic mean is meant. Other averaging procedures have also been suggested, yielding the $t$ index for the geometric mean and the $f$ index for the harmonic mean [8], thus giving more weight to highly cited papers. Another complementary index $m$ has been defined as the median of the number of citations in the $h$ core [9].

Hirsch [1] has already discussed the proportionality between the total number of citations and the square of the $h$ index. Several Hirsch-type indices have been proposed based correspondingly on the square root of the total number of citations to the papers in the core. Again the size of the core could be $h$ yielding once more a complementary index called $R$ [10]. Using the $g$ core reproduces the $g$ index according to Egghe's original definition [6]. For the definition of the $\hbar$ index [11] the total number of publications is used as core; for the weighted $h$ index $h_w$ [12] a subset of the $h$ core is taken into account. The $e$ index quantifies the square root of the excess citations to papers in the $h$ core [13].

Further variants which are not based on one of the above classifications have been suggested. E.g., the tapered $h$ index $h_T$ takes the citations to all publications into account in a complicated way [14]. The $\pi$ index depends on the total number of citations to papers in the so-called elite set [5]. Finally, the maxprod index was proposed to distinguish genies and hard workers from the typical researcher [15].

I have previously analyzed the citation records of 26 physicists from the Institute of Physics at the Chemnitz University of Technology and compared $h$, $g$, $A$, and $R$ [16]. It is the purpose of the present investigation to extend the previous discussion to the different variants mentioned above and their advantages and disadvantages. It will be shown that some of the proposed indices are not able to discriminate sufficiently between the different datasets. On the other hand, some of the indices lead to nearly the same ranking of the datasets which indicates a certain redundancy. It has even been suggested [9,17] that there are only two types of variants corresponding to the two standard bibliometric measures, either "number of publications" for the quantity or "total citation counts" for the impact [18]. In my point of view this division is exaggerated. But in any case it would certainly be prudent to take more than one or two indicators into consideration, especially for important evaluations.

Finally I would like to note that there are other types of $h$ index variants which are not considered in the present analysis, because they involve a manipulation of the raw data. For example in order to take the influence of self-citations into account one has to correct the citation counts which was done for the present 26 datasets [19]; for the consideration of multiple authors one can modify the paper counts which was also done for these datasets without [20] and with [21] self-citation corrections. It might also be reasonable to take the age of a paper into account so that citations to more recent publications get a higher weight [10]. Thus there is always room for further improvement.

**2 The citation records and indices based directly on the citation count**

The citation records of the 26 colleagues have been compiled from the Science Citation Index in the Web of Science in January and February 2007. In conformity with the previous analysis they are labelled A, B, C, ..., Z. Details of the determination of the datasets have been described elsewhere [19] in particular with respect to the precision problem, i.e. to establish that the considered publications have really been (co)authored by the investigated scientists and not by colleagues with the same name and the same initials and that on the other hand publications of the same person using different names (often due to marriage) are correctly collected [19].

The citation lists are arranged in decreasing order according to the number of citations $c(r)$, attributing the rank $r$ to the $r$th paper. (Although it is not relevant for the $h$ index, it is useful for definiteness to further sort papers with the same number of citations in, e.g., anti-chronological order thus specifying in particular, which papers are included in the core.) According to Hirsch's original definition the $h$ index can be easily read off this list as the largest rank for which [1]

$$h \leq c(h) \quad \text{while} \quad c(h+1) < h+1 . \tag{1}$$

Strictly speaking the second inequality is unnecessary, because it is implied in claiming that $h$ is the *largest* rank fulfilling Eq. (1).

For prominent scientists this definition yields relatively large values [1], i.e. a relatively large size of the $h$ core and correspondingly a significant precision problem. This has prompted the definition of the more restrictive $w$ index as the largest rank for which [3]

$$10w \leq c(w) \quad \text{while} \quad c(w+1) < 10(w+1) . \tag{2}$$

This often leads to very small values of the $w$ index which is supposed to be a measure of the number of "widely cited" papers. Somewhat higher values are usually obtained for the $h_2$ index defined as the largest rank for which [2]

$$h_2^2 \leq c(h_2) \quad \text{while} \quad c(h_2+1) < (h_2+1)^2 . \tag{3}$$

It should be noted that for clarity I have changed the notation writing $h_2$ instead of $h(2)$. From Eqs. (2,3) it is obvious that for $w < 10$ we have $w \leq h_2$, for $w = 10$ we get $h_2 = 10$, and for $w > 10$ we obtain $w \geq h_2$. For 25 of the here investigated datasets one finds $w < 10$, and in one case $w = 10 = h_2$. This means that the $w$ core is always smaller or equal to the $h_2$ core for these datasets. The respective values are presented in Table 1 together with the rank order in which the datasets appear when the list is sorted according the respective index. In Fig. 1 the citation records of six datasets are visualized and the determination of the indices is indicated. In this figure as well as in Table 1 the total number $n$ of publications as well as the number $n_1$ of publications which received at least one citation are also specified. Formally they can be determined in analogy to Eqs. (1,2,3). To be specific, $n$ is the largest rank for which

$$0 \leq c(n) \tag{4}$$

and $n_1$ is the largest rank for which

$$1 \leq c(n_1) \quad \text{while} \quad c(n_1+1) < 1. \tag{5}$$

The nearly equal total numbers of publications for the datasets G, J, P, R, U, and W motivated me to choose these 6 datasets for the visualization in Fig. 1.

It is worthwhile to note that the index values in Table 1 and Fig. 1 are given by the size of the corresponding core, i.e. the respective value of the rank on the horizontal axis in Fig. 1. Thus, although they give more or less preference to highly cited papers, the total number of citations in the respective core and thus the skewness is not explicitly taken into account in these indices.

It is not surprising that the Hirsch index leads to a significantly different ranking than the number $n$ of publications. Already the ranking in terms of the number of $n_1$ of cited publications clearly deviates from the ranking according to $n$. Strong differences occur in the rankings compared to the order according to $h$. The rank orders for $w$ and $h_2$ are not very different, although on first sight the rank order values in Table 1 give a different impression. But the large discrepancies between the ranks are misleading, because so many datasets are attributed the same rank due to the small integer values. (In the case of dataset U one more citation to the third paper would lead to an increase of the $w$ index, resulting in the new rank $O(w) = 14$ instead of 20.) On the other hand, the deviations to the $h$-index ranking are rather strong in some cases, specifically dataset G falls back in the $w$-index ranking as compared to $h$, while P and R advance significantly, as well as V. The difference between G and P is visualized in Fig. 1, where one can see that the dataset P starts with several highly cited papers, but then drops strongly while dataset G also starts relatively high, but the citation record stays high and in fact is the highest of the six datasets shown in Fig. 1 from $r = 7$ up to $r = n$.

The intention for the introduction of $h_2$ as well as for $w$ has mainly been to simplify the determination of the index values by reducing the size of the core. However, for the present datasets this has lead to so small numbers that the discrimination between different datasets is not possible anymore in many cases. This can be seen in Table 1 where many index values are equal or close to the median value 3.5 for $w$ and 5 for $h_2$. In order to quantify this problem I define a discrimination parameter $\Delta$ as the number of author pairs which cannot be distinguished because of coinciding index values: if an index appears twice in the list, this tie contributes 1 to this parameter, if an index value appears threefold, then there are 3 possible pairs so that the parameter increases by 3. For the $w$ index the value $w = 4$ appears tenfold, which means 45 possible pairings. This explains the high discrimination parameter $\Delta(w) = 71$ reflecting that $w$ is not a suitable index for the current analysis. Likewise $h_2$ does not allow a reasonable discrimination of most of the 26 datasets, one obtains $\Delta(h_2) = 63$. In conclusion, while $w$ and $h_2$ may have their merits for large citation records with many highly cited papers, for the average physicists with more moderate citation records they are too coarse.

In order to increase the discriminatory power of the $w$ index Wu [3] has introduced the additional factor $q$ for the least number of citations which are needed to increase the $w$ index from $w$ to $w+1$. This solves the discrimination problem, but in my opinion it remains doubtful, whether such a small number of papers as given by the $w$ index can be considered as representative for an average scientist's publication record. The same reservation applies to the $h_2$ index. On the other hand, $n$ is certainly a very crude indicator measuring only the productivity and not the impact. Likewise $n_1$ is strongly dominated by the productivity, because a single citation can be easily obtained, if necessary by self-citation. Thus the *h* index appears to be the best choice of the indicators in Table 1.

It is interesting to note that even for the *h* index five values appear more than once leading to a discrimination parameter $\Delta(h) = 14$. This is due to the fact that in spite of rather different citation records the *h*-index values cluster around the median value 14 for *h*. The problem of multiple index values can be significantly reduced by employing interpolated indices based on a piecewise linear interpolation of the rank-frequency function

$$\tilde{c}(x) = c(r) + (x-r)(c(r+1) - c(r)) \tag{6}$$

between *r* and *r*+1 and then defining an interpolated index $\tilde{h}$ by

$$\tilde{h} = \tilde{c}(\tilde{h}) . \tag{7}$$

(The tilde is used here and in the following to indicate the interpolated values.)[1] Then only two index value pairs remain, for $\tilde{h} = 13$ and $\tilde{h} = 14$. However, it is doubtful whether the resulting small differences between the index values which are now rational numbers are meaningful. For actual evaluation purposes this is certainly not the case.

**3 Index variants depending on the arithmetically averaged number of citations**

Although the indicators discussed in the previous section give more ($w, h_2$) or less ($n_1, n$) preference to highly cited manuscripts than *h* they are all based on a certain threshold with which the rank-frequency function is compared. Whether this result is self-consistently determined as for $w, h_2$, and *h* or whether it is fixed as for $n_1$ and *n*, it always means that once a publication has reached the respective core, further citations to it do not increase the index value. This is of course most extreme for *n*, for which no citation is necessary at all.

To remedy this situation one can utilize the average number of citations which is defined by

$$\bar{c}(r) = \frac{s(r)}{r} \tag{8}$$

in terms of the sum

$$s(r) = \sum_{r'=1}^{r} c(r') \tag{9}$$

of the number of citations to the papers up to rank *r*. In Fig. 2 the average citation counts are displayed for the same datasets which have been used for Fig. 1. Of course, due to the averaging the curves in Fig. 2 are much smoother than the corresponding curves in Fig. 1 and lie above those curves except for $r = 1$. Now the citation records can be more easily distinguished.

---

[1] The same interpolation was used [22, 23] for the definition of the so-called real variant $h_r$ of the *h* index.

Again the question arises how many papers should be taken into account for the definition of a suitable index. One solution is to utilize the $h$ core, i.e. to take the average number of citations to the papers up to rank $h$. The result

$$A = \bar{c}(h) = \frac{s(h)}{h} \qquad (10)$$

is called the $A$ index [4] and should obviously be used only in conjunction with $h$ itself.

Vinkler [5] defined the highly cited "elite set", the size of which is given by the square root of the total number of papers

$$n_\pi = \sqrt{n} \qquad (11)$$

(rounded to integer values), and discussed the respective average number of citations

$$\bar{c}(n_\pi) = \frac{s(n_\pi)}{n_\pi} \qquad . \qquad (12)$$

In the rank order in terms of this number as well as the $A$ index there are conspicuous changes, most notably scientists P and X advance, because they have one or several highly cited papers, before their citation records drop strongly.

In the present analysis the size of the "elite set" turned out to be always smaller than the $h$ core, so that the respective average from Eq. (12) is always larger than the $A$ index. The rankings are similar, with few exceptions, most notably for dataset D in one direction and datasets G and J in the other direction, which can be attributed to the particularly large number of publications for D and the relatively small numbers for G and J, influencing the size of the elite set. This points to an obvious disadvantage of this indicator, namely that it depends on the total number of publications and can therefore be easily influenced by inclusion or exclusion of irrelevant (meaning uncited) publications. To be specific, for dataset N the $n = 72$ publications define an elite set of 8 papers, but if this scientist publishes one more paper then $n = 73$ would yield an elite set of 9 publications and the average number of citations in the elite set would drop from 36 to 34.8. In fact, for 5 other datasets a similar effect could be observed: by adding two or three further publications the rounded values of $n_\pi$ would increase by one yielding a corresponding decrease of $\bar{c}(n_\pi)$. Moreover, it appears questionable in principle whether the total number of publications is a sensible quantity to define the size of the elite set, because spurious papers like editorials, comments, errata enhance the total number of publications inadequately. Also, at least in some fields conference proceedings which an author might be persuaded to contribute to, increase the number of publications but not the impact in terms of citations.

It is certainly more elegant to use a self-consistent definition which is possible in analogy to Eq. (1) by defining the *g* index [6] as the largest rank for which [7]

$$g \leq \bar{c}(g) \quad \text{while} \quad \bar{c}(g+1) < g+1. \tag{13}$$

Respective values of *g* and the other indices mentioned in this section are given in Table 2 and indicated in Fig. 2.

Also included in Table 2 are the values of the average number of citations to all papers as well as the highest citation counts $c_1$, because these can also be defined by Eq. (8) for a core size of $r = n$ and $r = 1$, respectively. Of course $c_1$ is a very crude indicator and will usually not be representative for a scientist's overall accomplishments. It is thus not surprising that the rank order for $c_1$ deviates strongly from all other rankings, most notably for dataset X which reflects the one-hit wonder on the one hand and dataset F, the enduring performer, on the other hand.

A disadvantage of $\bar{c}(n)$ is that it is necessary to collect all sources and items, i.e. to determine the number of citations to every paper. This means a considerable precision problem. Nevertheless $\bar{c}(n)$ has been proposed as "a superior indicator of scientific quality" [24], but as it is based on the total number of publications, my above criticism applies also in this case, namely that spurious papers and/or conference proceedings enhance *n* and thus reduce $\bar{c}(n)$ inadequately. I do not question the result of the analysis [24] that the average number of citations per paper is superior in terms of both accuracy and precision, but in my opinion it is not a suitable indicator for a scientist's achievements. As can be expected from the above discussion, the rank order in terms of $\bar{c}(n)$ significantly deviates from most other rank orders and the changes are in the opposite direction as compared to the changes in the rank order according to the total number *n* of publications: for example, dataset D comprises the largest number of publications and accordingly the average number of citations is extremely low. On the other hand, dataset G reaches position 3 in terms of the average number of citations, much better than its position in terms of all other indicators, and corresponding to a rather small number of papers and thus a large rank value on the *n*-sorted list.

Using the above mentioned interpolation (6) for the rank-frequency function one can determine a real-valued $\tilde{g}$ index from

$$\tilde{g} = \bar{c}(\tilde{g}) \tag{14}$$

in analogy to Eq. (7). This agrees with the linear interpolation of the sum (9) used [22, 23] for the definition of the real variant $g_r$ of the *g* index. However, this would mean a non-linear interpolation for the average number of citations. It is therefore easier to utilize a linear interpolation of the average number of citations in analogy to Eq. (6) for the definition of $\tilde{g}$ in Eq. (14). The corresponding graphical solution is visualized in Fig. 2. The difference between the two interpolation procedures is negligible, on average the $\tilde{g}$ values differ by 0.0025.

It is worthwhile to note that the index values displayed in Fig. 2 are determined by the average citation counts in the various core sets, i.e. in contrast to Fig. 1 the values are given by the value of the function $\bar{c}(r)$ and not the variable $r$. Therefore it is not surprising and in fact it is obvious from the definitions that for all the indicators in Table 2 further citations to the papers in the core set enhance the index values. For the discrete $g$ index a certain number of additional citations is necessary to have an effect, but for the interpolated version every additional citation makes an albeit small increase. A further advantage of the $\tilde{g}$ index is that the 8 ties which still occur for the $g$ index are resolved, so that the ranking according to $\tilde{g}$ becomes unique, as quantified by the discrimination parameter $\Delta$ in the last line of Table 2.

Comparing the rankings which result from the different indicators in Table 2 one observes even stronger rearrangements than in Table 1. As already mentioned, one extreme indicator, $c_1$, is most favourable for the one-hit wonder while the further indicators give more or less preference to the enduring performer. In the other extreme, $n$ quantifies the productivity, and when the size of the core depends on the (square root of the) total number of publications, then this number significantly influences the index value and thus the ranking according to $n$ can be expected to be more or less reflected in the ranking according to the $\pi$ index, as exemplified above. In conclusion, in my opinion the $g$ index or rather the $\tilde{g}$ index is the best choice among the indices and indicators discussed in this chapter.

**4 Index variants depending on other average citation numbers**

Instead of the average number of citations to the papers in the $h$ core as utilized for the definition of $A$, one could also employ the median number of citations to the papers in the $h$ core. The thus defined $m$ index [9] was introduced to measure the central tendency and thus deliberately disregards the skewness of the citation record within the $h$ core. I do not see any advantage in this procedure. Moreover, like the $A$ index, the $m$ index should only be used in conjunction with $h$ itself. Comparing the values of the $m$ index in Table 3 and the $A$ index in Table 2, we find that $m < A$ in all cases, which reflects the usually convex curvature (i.e. the usually non-negative second derivative) of the rank-frequency function. The ratio $m/A$ is around 0.8 in most cases. But there are some exceptions, most notably due to the relatively large values of $m$ for dataset Z but also for Q and U, and on the other hand due to the relatively small values of $m$ for datasets I and M. The largest difference in the rankings occurs for the one-hit wonder X.

For the definition for the average in Eq. (8) the arithmetic mean is utilized. Instead, one could also employ the harmonic mean

$$\bar{c}_{-1}(r) = \left( \frac{1}{r} \sum_{r'=1}^{r} c^{-1}(r') \right)^{-1} \tag{15}$$

for the definition of the $f$ index [8] as the largest rank for which

$$f \leq \bar{c}_{-1}(f) \quad \text{while} \quad \bar{c}_{-1}(f+1) < f+1 . \tag{16}$$

Another possibility is to utilize the geometric mean, i.e.

$$\bar{c}_{0}(r) = \left( \prod_{r'=1}^{r} c(r') \right)^{1/r} \tag{17}$$

which is equivalent to the logarithmic mean

$$\bar{c}_{0}(r) = \exp\left( \frac{1}{r} \sum_{r'=1}^{r} \ln c(r') \right) \tag{18}$$

used [8] for the definition of the $t$ index as the largest rank for which

$$t \leq \bar{c}_{0}(t) \quad \text{while} \quad \bar{c}_{0}(t+1) < t+1 . \tag{19}$$

Respective values are given in Table 3 and indicated in Fig. 2, too. In my opinion, the $f$ index as well as the $t$ index are as elegant as $h$ and $g$, because of their self-consistent definition. But their calculation is somewhat more involved and in contrast to the application in the field of economics [8] they do not have better discriminative power than the $g$ index in the present analysis, which can be seen in the last line of Table 3. (The particularly high value of $\Delta(t) = 15$ is due to the fivefold occurrence of $t = 20$.) This means that for the here investigated datasets the indices $f$ and $t$ are not more sensitive to small differences between the researchers than $g$. In conclusion, I consider the $g$ index to be the best choice of the indicators in Tables 2 and 3.

As described for the $g$ index, also $f$ and $t$ are enhanced by further citations to the papers in the core set, although a certain number of additional citations is necessary due to the discreteness. In principle one could also define interpolated values $\tilde{f}$ and $\tilde{t}$, although in these cases the interpolation is more involved. In order to enhance $m$, further citations to the papers exactly in the middle of the $h$ core are required. This shows that $m$ does not really fit into the class of index variants depending on the average number of citations. This is also noticeable in Fig. 2, where the respective data points are conspicuously separated from the citation record curves. Of course, also the values of the $t$ index and the $f$ index are not given by the curves in Fig. 2, because they are not determined by the arithmetic, but by the harmonic and geometric averaging.

## 5 Index variants depending on the square root of the summed number of citations

The minimum number of citations which is necessary to yield the $h$ index value $h$ is given by $h^2$, because $h$ papers with $h$ citations each are needed. This suggests to establish respective indices in terms of the square root of the summed number of citations which is displayed in Fig. 3 for the same datasets which have been used for Figs. 1 and 2. Hirsch [1] found empirically that the total number $s(n)$ of citations is proportional to $h^2$ with a proportionality constant $a$ of the order of 3 − 5, i.e.

$$h = \sqrt{s(n)/a} . \qquad (20)$$

The values of the proportionality constant $a$ in Table 4 confirm this observation for most of the present datasets. However, there are exceptions, datasets D and X have larger values of $a$, and datasets G, J, Q, and Y have values of $a \approx 2.5$. I could not identify a specific trend in the citation records, which might have produced these more extreme values. With the proportionality factor $a$ ranging between 2.37 and 5.41, on average $a = 3.53 \pm 0.75$, I conclude that the relation (20) is not well fulfilled.

Accordingly the $\hbar$ index which is defined as [11]

$$\hbar = \sqrt{s(n)/2} \qquad (21)$$

is only roughly proportional to $h$ with a proportionality constant around 1.3 as can be seen from Table 4. Again the above criticism applies, namely that it is difficult to establish the total citation count with high precision.

The constant $a$ quantifies the excess citations which are unnecessary for $h$. A value of $a = 1$ in Eq. (20) would reflect the minimum number of citations for $h$, which could be visualized as a flat citation record of $c(r) = h$ up to $r = h$ which, if plotted in Fig. 1, would yield a square with a corner on the diagonal straight line, see also Fig. 1 in [1]. A linearly decreasing citation record through the corner of this square would mean $a = 2$ and $\hbar = h$. This has been "assumed to be a lower bound quite generally" [1], because the second derivative of the rank-frequency function is usually nonnegative, what is more or less the case in the present analysis too, see Fig. 1. Indeed, $\hbar > h$ in all 26 cases, although there are 4 datasets (G, J, Q, Y) for which $\hbar/h \approx 1.1$ is rather small.

It is also interesting to see how much the citations to only the papers in the $h$ core go beyond the required minimum. This excess can be quantified by the $R$ index defined as [10]

$$R = \sqrt{s(h)} . \qquad (22)$$

According to Eq. (10) it is given by $R = \sqrt{Ah}$ and one obtains $R = h$ if all papers in the $h$ core have only received the minimum number $h$ of citations. In this case also $A = h$ and $m = h$.

Hirsch [1] has found that in the vast majority of cases the contributions to the total number of citations arise from the highly cited papers, i.e. $s(h) > s(n)/2$ what is equivalent with $R > \hbar$. This is confirmed in the present investigation where $R > \hbar$ in all cases except for dataset D with its extremely large number of publications.

A somewhat more complicated way of weighting the citations in the $h$ core was introduced by means of the weighted index [12]

$$h_w = \sqrt{s(r_0)} \qquad (23)$$

where $r_0 \leq h$, because it is determined as the largest rank $r_0$ for which

$$c(r_0) \geq \frac{s(r_0)}{h} \quad \text{while} \quad c(r_0+1) < \frac{s(r_0+1)}{h} \qquad . \qquad (24)$$

It is somewhat surprising that $r_0 = 6$ occurs 8 times. $r_0$ is small for high values of $c(1)$, especially for the one-hit wonder X one finds $r_0 = 1$. If $c(1)$ is relatively low, then $r_0$ is rather large, because a relatively flat rank-frequency function $c(r)$ yields a slowly increasing sum $s(r)$.

It should be noted that $h_w$ as well as $R$ are again supplementary indices, because they depened on the value of $h$. Although $r_0$ is often significantly smaller than $h$, often much less than $h/2$, the ranking in terms of $h_w$ is very similar to the ranking in terms of $R$. More surprising is that both rankings deviate only insignificantly from the ranking in terms of $\widetilde{g}$. Interestingly, even the ranking in terms of $\hbar$ yields only moderate changes.

Utilizing the square root of the summed citation counts, once more an elegant self-consistent definition is possible in analogy to Eq. (1) and Eq. (13) by defining the $g$ index as the largest rank for which [6]

$$g \leq \sqrt{s(g)} \quad \text{while} \quad \sqrt{s(g+1)} < g+1 \qquad . \qquad (25)$$

Taking the square and dividing by $g$ or $g+1$ one can easily see the equivalence with definition (13). Again an appropriate interpolation for the sum $s(g)$ or its square root allows us to define an interpolated $\widetilde{g}$ index in analogy to Eq. (7) and Eq. (14) as

$$\widetilde{g} = \sqrt{s(\widetilde{g})} \qquad . \qquad (26)$$

Respective values are visualized in Fig. 3. In my opinion the $\widetilde{g}$ index is the best choice of the indices in Table 4 because $h_w$ and $R$ are supplementary indices and $\hbar$ includes the complete tail of the rank-frequency function, i.e. also all the citations to all the lowly cited papers. One might argue that these are also part of a scientist's achievements, but they are certainly rather difficult to establish in view of the precision problem.

**6 Further index variants**

The π index is defined by the total number of citations in the above mentioned elite set scaled by an arbitrary prefactor [5]

$$\pi = 0.01 \; s(n_\pi) \qquad . \tag{27}$$

Thus it depends on the total number of publications via Eq. (11) so that the reservations which have been discussed with respect to Eq. (12) remain valid for Eq. (27). Moreover, the arbitrariness of the prefactor is a strange feature, which is probably just an attempt to scale the resulting values into a range comparable to the other indices. The π index is somewhat unique, because it is defined in terms of the summed number of citations rather than the square root of the sum or the average. However, taking the square root does not change the ranking and it is therefore not surprising that the π index leads to a ranking of the datasets similar to the $\tilde{g}$ ranking, see Table 5. The differences can be attributed mainly to the different sizes of the π core and the g core.

Different ways to quantify the citations to the papers in the h core have been given in the previous sections, namely the indices $A$, $h_w$, $m$, $R$, which are all complementary in one way or another to the h index. Another, unusual way to quantify the excess citations in the h core has recently been proposed [13]. The resulting e index which is defined as the square root of the excess citations

$$e = \sqrt{s(h) - h^2} \tag{28}$$

depends on h in contrast to the repeated claims of the author of Ref. [13], as is obvious from Eq. (28). It is closely related to A and R and is included in the present analysis for completeness. In my opinion it is not surprising that the ranking in terms of e is very similar to the A ranking.

An extraordinary way to weight the number of citations has been suggested [25] in analogy to the definition of the information entropy

$$S = -\sum_{r=1}^{n} \frac{c(r)}{s(n)} \log \frac{c(r)}{s(n)} \tag{29}$$

measuring the deviation from a uniform citation record, for which $c(r) = s(n)/n$ for all $r$ which would yield the maximum possible "entropy" $S_0 = \log n$. This is utilized for the definition of the s index [25]

$$s = \frac{1}{2} \sqrt{s(n) \frac{S}{S_0}} \qquad . \tag{30}$$

The proportionality to $\sqrt{s(n)}$ follows the expectation Eq. (20) by Hirsch [1] with a proportionality constant $a = 4$. The discussion in section 5 in connection to $\hbar$ already showed a variety of proportionality constants $a$ which is also reflected in the values of the s index in comparison with the h values, see Table 5. The comparison of s with $\sqrt{s(n)}$ shows a much better proportionality reflecting the observation that in the definition (30) the factor $\sqrt{S/S_0}$ = 0.896 ± 0.040 (where half of the standard deviation is caused by dataset X) does not lead to a good distinction between different datasets. Thus I conclude that the calculation of $S$ in Eq. (29) adds work, but not much insight.

In the previous sections the excess citations as well as all the citations to all the papers in the long tail of the publication records have been taken into account in the average number of citations $\bar{c}(n)$ and the $\hbar$ index. On the other hand as mentioned above Wu [3] has made an attempt to increase the discriminatory power of the $w$ index by determining the number of citations which are necessary to increase the $w$ index. A similar approach has been employed for the definition of the rational $h$ index [26] and the rational $g$ index [27] based on the number of further citations which are necessary to enhance $h$ or $g$. An interesting alternative appears to be the tapered index $h_T$ which attributes weights to all citations in such a way that $h_T = h$ again, if only the minimum number $h^2$ of citations has been received. All further citations not only to papers in the $h$ core, but also to all papers in the tail of the citation record are taken into account, but they are given a higher weight if they are more likely to enhance the $h$ index than the other citations. In practice this means that the ($h$+1)th citation to a paper in the $h$ core gets a higher weight than the ($h$+2)th citation. Likewise a citation to the ($h$+1)th paper gets a higher weight than a citation to the ($h$+2)th paper. To be specific the $i$th citation to the $r$th paper contributes with a weight

$$w(i,r) = \begin{cases} \dfrac{1}{2i-1} & \text{if } r \leq i \\ \dfrac{1}{2r-1} & \text{if } r \geq i \end{cases} \quad (31)$$

to the tapered $h$ index [12]

$$h_T = \sum_{r=1}^{n} \sum_{i=1}^{c(r)} w(i,r) \quad . \quad (32)$$

In this way on the one hand the strongly skewed citation record with usually a few highly cited papers is considered in a tapered way, on the other hand a similar tapering is applied to the long tail of the citation record. The first $h$ citations to the first $h$ papers contribute exactly with the weight $h$ to $h_T$. A disadvantage is that $h_T$ is somewhat more difficult to calculate than the other indices and that it depends on the complete citation record which is difficult to establish with the necessary accuracy. Therefore, although I consider the idea intriguing, in my opinion this index is unlikely to be utilized frequently. Moreover, the values of $h_T$ which are given in Table 5 are not significantly different from $0.9\sqrt{s(n)}$ so that its determination does not seem to be worth the extra effort. (Again the standard deviation of the prefactor 0.888 ± 0.050 is mainly caused by dataset X.)

Finally a rather exotic variant is determined as the maximum of the product of rank and citation frequency [15]

$$x = \max_r (r\, c(r)) \quad , \tag{33}$$

hence the original name Maxprod. For simplicity of the notation I utilize here the label $x$ because this symbol is often used instead of a multiplication sign. Usually $x > h^2$, while $x = h^2$ if there are only exactly $h$ citations to each of the $h$ papers in the $h$ core. The rank $r_x$ for which the product $r_x\, c(r_x)$ is maximum can be smaller than, equal to, or larger than the $h$ index, depending on the specific rank-frequency distribution. Thus the $x$ index is intended [15] to discriminate between genies, typical scientists, and hard workers, respectively, in terms of the rank $r_x$. The index value $x$ favours enduring performers rather than one-hit-wonders. Looking at the results of the present investigation in Table 5, one can observe that the $r_x$ values do indeed distinguish the datasets in a distinct way. Whether this reflects the mentioned three categories remains a matter of interpretation. The values of the $x$ index are of course rather large, but the ranking in terms of $x$ is not much different from the $\widetilde{g}$ ranking, so that once more the question arises whether the calculation of this index is worth the extra effort.

## 7 Correlation coefficients

The above made observations concerning the rankings according to the different indices and indicators can be quantified by calculating Spearman's rank-order correlation coefficients $\kappa$. Respective values are presented in Table 6. To avoid misunderstandings, I point out that for the calculation of these coefficients one should not take the rank values from Tables 1−5, but rather correct multiple entries to the respective intermediate values. This means that for example the tenfold value $o(w) = 4$ should be replaced by the value 8.5 and the value $o(w) = 25$ appearing twice needs to be replaced by 25.5.

In previous publications often Pearson's correlation coefficients have been utilized. The respected values are also given in Table 6. However, they are only meaningful, if the values are approximately distributed according to the normal distribution. As previously shown [21], this is the case for the Hirsch index and some of its variants for the here investigated 26 datasets. But for some of the indices and indicators in the present investigation this is not true. Therefore the values of Pearson's correlation coefficients in Table 6 should be taken with caution and in the following I discuss mostly the rank-order correlation coefficients.

Recalling the above observations, it is not surprising that the number $n$ of publications and the number $n_1$ of cited publications correlate only weakly with most of the other indices and indicators, with a value as low as $\kappa(n,\bar{c}(n)) = 0.066$. A closer inspection of Table 6 shows that there are 12 values of $\kappa$ below $\kappa(c_1,m) = 0.556$ which is the smallest correlation remaining if $n$ and $n_1$ are not taken into account. Next, the average number of citations $\bar{c}(n)$ to all publications and the highest number of citations $c_1$ are not so strongly correlated with the remaining indices and 20 more values of $\kappa$ up to 0.75 drop out of the table if these indicators are also not taken into consideration. The smallest remaining correlation is now $\kappa(s,\bar{c}(\sqrt{n})) = 0.753$. It suggests itself to exclude $\bar{c}(\sqrt{n})$ next, avoiding 5 further $\kappa$ values below 0.78 with $\kappa(w,m) = 0.779$ the smallest value of the rest.

Among the remaining indices now $w$, $A$, and $e$ show the weakest correlations, although all the values are already rather high. Nevertheless, excluding $w$, $A$, and $e$ means that 16 $\kappa$ values below 0.85 drop out of the table and the minimum is now given by $\kappa(m,s) = 0.847$. Now the $m$ index shows the smallest correlation coefficient with all other remaining indices and leaving it out means that the 13 $\kappa$ values of $m$ with the other indices which are all below 0.895 fall out of the table and $\kappa(\pi,h) = 0.897$ is the remaining minimum.

Finally, if one also deletes $h_2$ and $\pi$, further eight $\kappa$ values below 0.93 vanish and the minimum is now at $\kappa(R,s) = 0.935$. This clearly demonstrates that at least the remaining indices $f$, $g$, $\tilde{g}$, $h$, $h_w$, $h_T$, $\hbar$, $R$, $s$, $t$, and $x$ yield very similar rankings. It is interesting to note that Pearson's correlation coefficients between the remaining 10 indices (or 11 indices, if one counts $g$ and $\tilde{g}$ separately) are all higher than 0.95, thus also demonstrating the very high correlation between these indices.

## 8 Concluding remarks

My analysis is based on the citation records of 26 scientists. This is a rather small sample. Recently significantly larger samples have been investigated with respect to the $h$ index, e.g. the citation data of 588 Greek professors [28], 396 material scientists from Mexico, Chile, and Columbia [29], and 402 members of the Brazilian Academy of Sciences [30]. But it has not been my aim to increase the number of datasets compared to my former studies but rather to extend the number of index variants which comprised only $h$, $A$, $R$, and $g$ in my previous analysis [16].

For this purpose twenty different variants of the Hirsch index and comparable indicators which give more or less preference to highly cited papers have been compared. Another comprehensive review has lately been given by Egghe [31]. I have tried to include into the present analysis all index variants which are directly based on the number of citations and do not require any manipulation of the raw citation frequency data, like excluding self-citations, or fractionalized counting of multiauthor papers, or aging effects. It was shown that many indices are highly correlated, while low correlations occur with indices and indicators based on the total number of publications or the number of citations to a core set based in one way or another on the total number of papers.

Among the highly correlated indices I favour the $\tilde{g}$ variant, because it is not a complementary index requiring first the determination of $h$, but rather follows from a self-consistent definition. If the assessment of the interpolated variant $\tilde{g}$ appears too difficult, one might as well utilize the simpler version $g$, which can be established from the citation record nearly as easily as the $h$ index. Of course it would be helpful, if one could convince the Thomson Reuters to provide in the ISI Web of Science not only the number of citations but also the average number of citations up to a given rank, when the citation record is sorted according to the times cited. In comparison to the other averages the arithmetic average is simple, and therefore the $g$ index is easier to calculate than $f$ and $t$. It is slightly more difficult to determine $g$ than $h$, but I think it is worth the small additional effort because highly cited papers are given additional weight. Therefore in my opinion it is fairer than $h$ and, what is more, the observed changes in the ranking yield significant differences, as quantified in Table 3. Of course, my preference is a matter of taste and there are now so many variants, that different scientists are likely to favour different variants, especially if they are treated more kindly by one variant than by another. In this context I would like to point out that my choice of $\tilde{g}$ as the preferable variant is not biased by my obtained rank, because I end up with the same rank for nearly all the indices and indicators and with exactly the same rank for all the highly correlated indices determined in section 7.

Whatever index one chooses to evaluate, one should always keep in mind that the quality of the database is decisive. Although the "distinct author" feature has now been introduced in the Web of Science, it remains a formidable task to establish the citation data of an individual scientist with high accuracy. The precision problem is often underestimated in actual applications. That is one reason why the discussion about the usefulness of these rankings is ongoing. Of course, administrators and other bureaucrats like them or love them. Scientists are more skeptical, not only if they do not end up on high positions in the ranking. But I would like to point out that this discussion of the citation impact approach is not new. It has recently come to my attention that citation analysis has already been performed nearly a century ago [32] even then with the purpose of allocating or not allocating funds.

**Table 1** Indices and indicators based on the number of citations for different core sizes as defined in Eqs. (1-5, 7). The datasets are sorted according to the interpolated Hirsch index $\tilde{h}$. The rank order for the other indices and indicators is denoted by $O$(index). The last line of the table shows the number of author pairs which cannot be distinguished because of coinciding index values.

| data set | $w$ | $h_2$ | $h$ | $\tilde{h}$ | $n_1$ | $n$ | $O(w)$ | $O(h_2)$ | $O(h)$ | $O(n_1)$ | $O(n)$ |
|---|---|---|---|---|---|---|---|---|---|---|---|
| A | 10 | 10 | 39 | 39.0 | 250 | 290 | 1 | 1 | 1 | 2 | 2 |
| B | 7 | 8 | 27 | 27.5 | 214 | 270 | 2 | 2 | 2 | 3 | 3 |
| C | 5 | 7 | 23 | 23.0 | 103 | 126 | 3 | 3 | 3 | 5 | 5 |
| D | 4 | 6 | 20 | 20.0 | 259 | 322 | 4 | 4 | 4 | 1 | 1 |
| E | 4 | 6 | 19 | 19.3 | 57 | 63 | 4 | 4 | 5 | 11 | 15 |
| F | 4 | 5 | 18 | 18.0 | 107 | 131 | 4 | 8 | 6 | 4 | 4 |
| G | 3 | 5 | 17 | 17.0 | 47 | 49 | 14 | 8 | 7 | 17 | 20 |
| H | 4 | 6 | 16 | 16.0 | 47 | 70 | 4 | 4 | 8 | 17 | 12 |
| I | 4 | 6 | 15 | 15.3 | 53 | 65 | 4 | 4 | 9 | 14 | 14 |
| J | 4 | 5 | 15 | 15.0 | 32 | 51 | 4 | 8 | 9 | 23 | 19 |
| K | 3 | 5 | 14 | 14.5 | 56 | 79 | 14 | 8 | 11 | 12 | 8 |
| L | 4 | 5 | 14 | 14.4 | 67 | 88 | 4 | 8 | 11 | 6 | 6 |
| M | 4 | 5 | 14 | 14.0 | 60 | 70 | 4 | 8 | 11 | 9 | 12 |
| N | 3 | 5 | 14 | 14.0 | 61 | 72 | 14 | 8 | 11 | 8 | 11 |
| O | 3 | 4 | 13 | 13.3 | 66 | 77 | 14 | 17 | 15 | 7 | 10 |
| P | 4 | 5 | 13 | 13.0 | 37 | 47 | 4 | 8 | 15 | 19 | 21 |
| Q | 2 | 4 | 13 | 13.0 | 59 | 86 | 20 | 17 | 15 | 10 | 7 |
| R | 4 | 5 | 12 | 12.3 | 37 | 46 | 4 | 8 | 18 | 19 | 22 |
| S | 3 | 4 | 12 | 12.0 | 48 | 61 | 14 | 17 | 18 | 16 | 16 |
| T | 2 | 4 | 10 | 10.7 | 56 | 78 | 20 | 17 | 20 | 12 | 9 |
| U | 2 | 4 | 10 | 10.5 | 34 | 44 | 20 | 17 | 20 | 22 | 23 |
| V | 3 | 4 | 10 | 10.3 | 49 | 60 | 14 | 17 | 20 | 15 | 17 |
| W | 2 | 3 | 9 | 9.0 | 37 | 53 | 20 | 23 | 23 | 19 | 18 |
| X | 1 | 3 | 8 | 8.0 | 29 | 35 | 25 | 23 | 24 | 24 | 24 |
| Y | 1 | 3 | 7 | 7.0 | 19 | 25 | 25 | 23 | 25 | 25 | 25 |
| Z | 2 | 3 | 5 | 5.3 | 12 | 15 | 20 | 23 | 26 | 26 | 26 |
| Δ | 71 | 63 | 14 | 2 | 5 | 1 | | | | | |

**Table 2** Indices and indicators based on the arithmetically averaged number of citations for different core sizes as defined in Eqs. (10, 12-14). The highest number of citations $c_1 = \bar{c}(1)$ and the average number of citations $\bar{c}(n)$ to all publications are also given. The datasets are sorted according to $\tilde{g}$. The rank order for the other indices and indicators is denoted by $\mathcal{O}$(index). The last line of the table shows the number of author pairs which cannot be distinguished because of coinciding index values.

| data set | $c_1$ | $\bar{c}(n_\pi)$ | $A$ | $g$ | $\tilde{g}$ | $\bar{c}(n)$ | $\mathcal{O}(c_1)$ | $\mathcal{O}(\bar{c}(n_\pi))$ | $\mathcal{O}(A)$ | $\mathcal{O}(g)$ | $\mathcal{O}(\bar{c}(n))$ |
|---|---|---|---|---|---|---|---|---|---|---|---|
| A | 457 | 144.1 | 93.9 | 67 | 67.1 | 20.7 | 1 | 1 | 1 | 1 | 2 |
| B | 182 | 83.5 | 62.6 | 45 | 45.6 | 11.8 | 4 | 3 | 2 | 2 | 7 |
| E | 279 | 109.5 | 62.4 | 37 | 37.2 | 22.8 | 2 | 2 | 3 | 3 | 1 |
| C | 129 | 66.8 | 47.3 | 36 | 36.7 | 13.2 | 6 | 5 | 4 | 4 | 6 |
| D | 73 | 37.2 | 35.5 | 29 | 29.8 | 6.6 | 11 | 14 | 8 | 5 | 21 |
| I | 149 | 68.8 | 46.1 | 28 | 28.8 | 13.6 | 5 | 4 | 5 | 6 | 4 |
| F | 53 | 39.1 | 32.2 | 26 | 26.6 | 8.6 | 17 | 12 | 11 | 7 | 14 |
| H | 70 | 50.0 | 35.9 | 26 | 26.2 | 10.7 | 12 | 7 | 7 | 7 | 9 |
| P | 108 | 63.7 | 41.5 | 24 | 24.7 | 13.4 | 8 | 6 | 6 | 9 | 5 |
| M | 100 | 47.1 | 34.0 | 24 | 24.1 | 10.4 | 9 | 8 | 10 | 9 | 10 |
| G | 57 | 40.3 | 28.4 | 23 | 23.9 | 14.2 | 14 | 11 | 14 | 11 | 3 |
| J | 112 | 46.9 | 32.1 | 23 | 23.6 | 11.3 | 7 | 9 | 12 | 11 | 8 |
| L | 64 | 37.4 | 30.6 | 22 | 22.7 | 7.7 | 13 | 13 | 13 | 13 | 16 |
| N | 55 | 36.0 | 27.7 | 22 | 22.1 | 9.5 | 15 | 16 | 15 | 13 | 13 |
| K | 55 | 33.6 | 27.7 | 21 | 22.0 | 7.5 | 15 | 17 | 15 | 15 | 17 |
| R | 53 | 36.3 | 27.0 | 19 | 19.8 | 9.8 | 17 | 15 | 17 | 16 | 12 |
| O | 47 | 26.3 | 22.8 | 19 | 19.1 | 7.1 | 19 | 21 | 20 | 16 | 19 |
| S | 40 | 27.9 | 22.8 | 18 | 18.2 | 7.2 | 22 | 19 | 21 | 18 | 18 |
| X | 204 | 44.0 | 35.1 | 18 | 18.2 | 9.9 | 3 | 10 | 9 | 18 | 11 |
| V | 79 | 27.8 | 24.4 | 17 | 17.2 | 6.5 | 10 | 20 | 18 | 20 | 22 |
| U | 41 | 28.3 | 23.7 | 17 | 17.2 | 8.0 | 21 | 18 | 19 | 20 | 15 |
| Q | 24 | 18.9 | 17.1 | 15 | 15.9 | 4.9 | 25 | 23 | 23 | 22 | 24 |
| T | 31 | 18.7 | 18.0 | 15 | 15.1 | 4.8 | 23 | 24 | 22 | 22 | 25 |
| W | 42 | 17.4 | 15.6 | 13 | 13.2 | 4.9 | 20 | 25 | 25 | 24 | 23 |
| Z | 25 | 19.8 | 17.0 | 10 | 10.0 | 6.9 | 24 | 22 | 24 | 25 | 20 |
| Y | 19 | 12.6 | 11.0 | 9 | 9.5 | 4.6 | 26 | 26 | 26 | 26 | 26 |
| Δ | 2 | 0 | 1 | 8 | 0 | 0 | | | | | |

**Table 3** Indices based on other average number of citations for different core sizes as defined in Eqs. (16, 19). The median number *m* of citations to papers in the *h* core is also given. The values of *h* and *g* are repeated for easier comparison. The datasets are sorted according to $\tilde{g}$. The rank order for the indices is denoted by $O$(index). The last line of the table shows the number of author pairs which cannot be distinguished because of coinciding index values.

| data set | h | f | t | g | m | $O(h)$ | $O(f)$ | $O(t)$ | $O(g)$ | $O(m)$ |
|---|---|---|---|---|---|---|---|---|---|---|
| A | 39 | 53 | 58 | 67 | 72.0 | 1 | 1 | 1 | 1 | 1 |
| B | 27 | 36 | 40 | 45 | 47.0 | 2 | 2 | 2 | 2 | 2 |
| E | 19 | 25 | 28 | 37 | 38.0 | 5 | 5 | 4 | 3 | 4 |
| C | 23 | 31 | 33 | 36 | 40.0 | 3 | 3 | 3 | 4 | 3 |
| D | 20 | 26 | 27 | 29 | 30.5 | 4 | 4 | 5 | 5 | 5 |
| I | 15 | 20 | 23 | 28 | 24.0 | 9 | 9 | 7 | 6 | 12 |
| F | 18 | 23 | 24 | 26 | 29.0 | 6 | 6 | 6 | 7 | 7 |
| H | 16 | 21 | 23 | 26 | 30.5 | 8 | 7 | 7 | 7 | 5 |
| P | 13 | 16 | 19 | 24 | 27.0 | 15 | 15 | 15 | 9 | 8 |
| M | 14 | 18 | 20 | 24 | 21.0 | 11 | 12 | 10 | 9 | 17 |
| G | 17 | 21 | 22 | 23 | 26.0 | 7 | 7 | 9 | 11 | 10 |
| J | 15 | 19 | 20 | 23 | 23.0 | 9 | 10 | 10 | 11 | 14 |
| L | 14 | 18 | 20 | 22 | 23.0 | 11 | 12 | 10 | 13 | 14 |
| N | 14 | 18 | 20 | 22 | 26.0 | 11 | 12 | 10 | 13 | 10 |
| K | 14 | 19 | 20 | 21 | 26.5 | 11 | 10 | 10 | 15 | 9 |
| R | 12 | 15 | 17 | 19 | 19.5 | 18 | 17 | 16 | 16 | 18 |
| O | 13 | 16 | 17 | 19 | 18.0 | 15 | 15 | 16 | 16 | 19 |
| S | 12 | 15 | 16 | 18 | 18.0 | 18 | 17 | 18 | 18 | 19 |
| X | 8 | 10 | 11 | 18 | 10.5 | 24 | 24 | 23 | 18 | 25 |
| V | 10 | 13 | 14 | 17 | 14.5 | 20 | 21 | 21 | 20 | 23 |
| U | 10 | 14 | 15 | 17 | 23.5 | 20 | 20 | 19 | 20 | 13 |
| Q | 13 | 15 | 15 | 15 | 17.0 | 15 | 17 | 19 | 22 | 21 |
| T | 10 | 13 | 14 | 15 | 15.5 | 20 | 21 | 21 | 22 | 22 |
| W | 9 | 11 | 11 | 13 | 12.0 | 23 | 23 | 23 | 24 | 24 |
| Z | 5 | 6 | 8 | 10 | 23.0 | 26 | 26 | 26 | 25 | 14 |
| Y | 7 | 8 | 9 | 9 | 10.0 | 25 | 25 | 25 | 26 | 26 |
| Δ | 14 | 10 | 15 | 8 | 6 | | | | | |

**Table 4** Indices based on the square root of the summed number of citations for different core sizes and defined in Eqs. (21-23, 25, 26). The rank $r_0$ which is needed for the calculation of $h_w$ (see Eq. (24)) and the proportionality constant $a$ from Eq. (20) are also given. The datasets are sorted according to $\tilde{g}$. The rank order for the other indices is denoted by $O$(index). The last line of the table shows the number of author pairs which cannot be distinguished because of coinciding index values.

| data set | $h_w$ | $r_0$ | $R$ | $\hbar$ | $a$ | $g$ | $\tilde{g}$ | $O(h_w)$ | $O(R)$ | $O(\hbar)$ | $O(g)$ |
|---|---|---|---|---|---|---|---|---|---|---|---|
| A | 51.7 | 20 | 60.5 | 54.8 | 3.94 | 67 | 67.1 | 1 | 1 | 1 | 1 |
| B | 35.3 | 14 | 41.1 | 39.9 | 4.36 | 45 | 45.6 | 2 | 2 | 2 | 2 |
| E | 28.2 | 6 | 34.4 | 26.8 | 3.99 | 37 | 37.2 | 4 | 3 | 5 | 3 |
| C | 28.5 | 13 | 33.0 | 28.8 | 3.14 | 36 | 36.7 | 3 | 4 | 4 | 4 |
| D | 23.6 | 13 | 26.6 | 32.6 | 5.31 | 29 | 29.8 | 5 | 5 | 3 | 5 |
| I | 22.3 | 6 | 26.3 | 21.0 | 3.93 | 28 | 28.8 | 6 | 6 | 7 | 6 |
| F | 20.7 | 11 | 24.1 | 23.7 | 3.48 | 26 | 26.6 | 8 | 7 | 6 | 7 |
| H | 21.4 | 10 | 24.0 | 19.4 | 2.93 | 26 | 26.2 | 7 | 8 | 8 | 7 |
| P | 20.5 | 6 | 23.2 | 17.8 | 3.73 | 24 | 24.7 | 9 | 9 | 13 | 9 |
| M | 18.3 | 6 | 21.8 | 19.1 | 3.70 | 24 | 24.1 | 11 | 12 | 9 | 9 |
| G | 18.3 | 9 | 22.0 | 18.7 | 2.41 | 23 | 23.9 | 10 | 10 | 10 | 11 |
| J | 18.1 | 7 | 21.9 | 16.9 | 2.55 | 23 | 23.6 | 12 | 11 | 15 | 11 |
| L | 17.8 | 8 | 20.7 | 18.5 | 3.47 | 22 | 22.7 | 13 | 13 | 12 | 13 |
| N | 17.7 | 9 | 19.7 | 18.5 | 3.51 | 22 | 22.1 | 14 | 14 | 11 | 13 |
| K | 16.8 | 8 | 19.7 | 17.3 | 3.04 | 21 | 22.0 | 15 | 14 | 14 | 15 |
| R | 15.4 | 6 | 18.0 | 15.0 | 3.13 | 19 | 19.8 | 16 | 16 | 17 | 16 |
| O | 14.9 | 8 | 17.2 | 16.6 | 3.25 | 19 | 19.1 | 17 | 17 | 16 | 16 |
| S | 13.8 | 6 | 16.6 | 14.8 | 3.05 | 18 | 18.2 | 19 | 19 | 18 | 18 |
| X | 14.3 | 1 | 16.8 | 13.2 | 5.41 | 18 | 18.2 | 18 | 18 | 23 | 18 |
| V | 13.0 | 4 | 15.6 | 13.9 | 3.89 | 17 | 17.2 | 21 | 20 | 20 | 20 |
| U | 13.4 | 6 | 15.4 | 13.2 | 3.51 | 17 | 17.2 | 20 | 21 | 22 | 20 |
| Q | 13.0 | 9 | 14.9 | 14.5 | 2.50 | 15 | 15.9 | 21 | 22 | 19 | 22 |
| T | 11.4 | 6 | 13.4 | 13.7 | 3.75 | 15 | 15.1 | 23 | 23 | 21 | 22 |
| W | 10.1 | 5 | 11.8 | 11.4 | 3.22 | 13 | 13.2 | 24 | 24 | 24 | 24 |
| Z | 8.5 | 3 | 9.2 | 7.2 | 4.12 | 10 | 10.0 | 25 | 25 | 26 | 25 |
| Y | 7.9 | 5 | 8.8 | 7.6 | 2.37 | 9 | 9.5 | 26 | 26 | 25 | 26 |
| Δ | 1 | | 1 | 0 | | 8 | 0 | | | | |

**Table 5** Further indices defined in Eqs. (27, 28, 30, 32, 33). The ratios $s/\sqrt{s(n)}$ and $h_T/\sqrt{s(n)}$ and the rank $r_x$ which is needed for the calculation of $x$ are also given. The datasets are sorted according to $\tilde{g}$. The rank order for the indices is denoted by $\mathcal{O}(index)$. The last line of the table shows the number of author pairs which cannot be distinguished because of coinciding index values.

| data set | $\pi$ | $e$ | $s$ | $\dfrac{s}{\sqrt{s(n)}}$ | $h_T$ | $\dfrac{h_T}{\sqrt{s(n)}}$ | $x$ | $r_x$ | $\mathcal{O}(\pi)$ | $\mathcal{O}(e)$ | $\mathcal{O}(s)$ | $\mathcal{O}(h_T)$ | $\mathcal{O}(x)$ |
|---|---|---|---|---|---|---|---|---|---|---|---|---|---|
| A | 24.5 | 46.3 | 35.3 | 0.46 | 67.6 | 0.87 | 1665 | 45 | 1 | 1 | 1 | 1 | 1 |
| B | 13.4 | 31.0 | 25.7 | 0.46 | 48.5 | 0.86 | 938 | 67 | 2 | 2 | 2 | 2 | 2 |
| E | 8.8 | 28.7 | 16.4 | 0.43 | 31.2 | 0.82 | 522 | 2 | 3 | 3 | 5 | 5 | 5 |
| C | 7.4 | 23.6 | 18.5 | 0.45 | 37.4 | 0.92 | 609 | 29 | 4 | 4 | 4 | 4 | 4 |
| D | 6.7 | 17.6 | 21.5 | 0.47 | 37.5 | 0.81 | 744 | 93 | 5 | 8 | 3 | 3 | 3 |
| I | 5.5 | 21.6 | 12.9 | 0.43 | 25.4 | 0.85 | 284 | 2 | 6 | 5 | 7 | 7 | 9 |
| F | 4.3 | 16.0 | 15.6 | 0.46 | 30.2 | 0.90 | 408 | 34 | 8 | 11 | 6 | 6 | 6 |
| H | 4.0 | 17.8 | 12.1 | 0.44 | 25.3 | 0.92 | 294 | 6 | 9 | 7 | 11 | 8 | 7 |
| P | 4.5 | 19.2 | 10.7 | 0.43 | 21.1 | 0.84 | 245 | 5 | 7 | 6 | 15 | 16 | 12 |
| M | 3.8 | 16.7 | 12.2 | 0.45 | 23.7 | 0.88 | 221 | 17 | 10 | 9 | 10 | 11 | 15 |
| G | 2.8 | 13.9 | 12.5 | 0.47 | 24.9 | 0.94 | 289 | 17 | 15 | 14 | 8 | 9 | 8 |
| J | 3.3 | 16.0 | 10.4 | 0.43 | 21.9 | 0.91 | 247 | 13 | 12 | 10 | 16 | 14 | 11 |
| L | 3.4 | 15.2 | 11.8 | 0.45 | 23.7 | 0.91 | 234 | 13 | 11 | 12 | 12 | 11 | 13 |
| N | 2.9 | 13.9 | 12.2 | 0.47 | 23.9 | 0.91 | 261 | 29 | 14 | 15 | 9 | 10 | 10 |
| K | 3.0 | 13.9 | 10.9 | 0.45 | 22.8 | 0.93 | 228 | 19 | 13 | 15 | 13 | 13 | 14 |
| R | 2.5 | 13.4 | 9.6 | 0.45 | 19.4 | 0.91 | 160 | 20 | 17 | 17 | 17 | 17 | 20 |
| O | 2.4 | 11.3 | 10.9 | 0.46 | 21.3 | 0.91 | 203 | 29 | 18 | 21 | 14 | 15 | 17 |
| S | 2.2 | 11.4 | 9.6 | 0.46 | 19.2 | 0.92 | 160 | 16 | 19 | 20 | 18 | 18 | 20 |
| X | 2.6 | 14.7 | 6.8 | 0.37 | 13.2 | 0.71 | 204 | 1 | 16 | 13 | 24 | 24 | 16 |
| V | 2.2 | 12.0 | 8.8 | 0.44 | 17.1 | 0.87 | 116 | 29 | 20 | 18 | 21 | 22 | 23 |
| U | 2.0 | 11.7 | 8.5 | 0.45 | 17.4 | 0.93 | 138 | 6 | 21 | 19 | 22 | 21 | 22 |
| Q | 1.7 | 7.3 | 9.4 | 0.46 | 19.0 | 0.92 | 189 | 21 | 22 | 25 | 19 | 19 | 18 |
| T | 1.7 | 8.9 | 8.9 | 0.46 | 17.5 | 0.90 | 162 | 18 | 23 | 22 | 20 | 20 | 19 |
| W | 1.2 | 7.7 | 7.3 | 0.45 | 14.5 | 0.90 | 112 | 28 | 24 | 24 | 23 | 23 | 24 |
| Z | 0.8 | 7.7 | 4.4 | 0.43 | 8.8 | 0.87 | 69 | 3 | 25 | 23 | 26 | 26 | 25 |
| Y | 0.6 | 5.3 | 4.9 | 0.46 | 10.2 | 0.95 | 52 | 13 | 26 | 26 | 25 | 25 | 26 |
| Δ | 0 | 1 | 0 | | 1 | | 1 | | | | | | |

**Table 6** Correlation coefficients between the various indices and indicators as specified on the diagonal of the table. Values for Spearman's rank-order correlation coefficients are given in the upper right triangle, values for Pearson's correlation coefficients are presented in the lower left triangle. For a better readability, values are given in percent. Lines separate groups of indices and indicators introduced in Sects. 2−6 and appearing in Tables 1−5.

| | | | | | | | | | | | | | | | | | | | | | | | | |
|---|---|---|---|---|---|---|---|---|---|---|---|---|---|---|---|---|---|---|---|---|---|---|---|---|
| $w$ | 92 | 85 | 60 | 56 | 69 | 82 | 82 | 90 | 91 | 68 | 85 | 88 | 78 | 90 | 90 | 88 | 90 | 87 | 84 | 86 | 79 |
| 95 | $h_2$ | 93 | 64 | 62 | 70 | 84 | 87 | 95 | 95 | 72 | 94 | 96 | 88 | 95 | 95 | 95 | 93 | 90 | 93 | 95 | 91 |
| 94 | 96 | $h$ | 73 | 71 | 63 | 77 | 80 | 94 | 94 | 66 | 100 | 99 | 86 | 94 | 94 | 97 | 90 | 83 | 97 | 98 | 96 |
| 75 | 76 | 83 | $n_1$ | 95 | 32 | 35 | 47 | 63 | 64 | 15 | 72 | 71 | 55 | 62 | 61 | 76 | 63 | 47 | 80 | 77 | 68 |
| 72 | 75 | 80 | 100 | $n$ | 25 | 29 | 43 | 58 | 58 | 7 | 70 | 68 | 53 | 57 | 57 | 71 | 58 | 43 | 74 | 73 | 67 |
| 74 | 73 | 76 | 53 | 49 | $c_1$ | 90 | 91 | 80 | 79 | 79 | 63 | 67 | 56 | 79 | 80 | 68 | 83 | 90 | 62 | 63 | 72 |
| 84 | 86 | 84 | 55 | 51 | 94 | $\bar{c}(n_\pi)$ | 97 | 91 | 91 | 93 | 78 | 82 | 75 | 92 | 91 | 81 | 92 | 97 | 75 | 77 | 83 |
| 89 | 91 | 89 | 66 | 63 | 93 | 99 | $A$ | 94 | 94 | 84 | 81 | 85 | 79 | 94 | 94 | 86 | 96 | 99 | 80 | 82 | 88 |
| 94 | 96 | 97 | 77 | 74 | 87 | 94 | 97 | $g$ | 100 | 79 | 94 | 96 | 87 | 100 | 100 | 97 | 99 | 96 | 94 | 95 | 95 |
| 94 | 96 | 97 | 77 | 74 | 86 | 93 | 97 | 100 | $\tilde{g}$ | 79 | 95 | 97 | 87 | 100 | 100 | 97 | 99 | 95 | 94 | 95 | 95 |
| 66 | 71 | 67 | 29 | 24 | 83 | 92 | 86 | 77 | 77 | $\bar{c}(n)$ | 67 | 71 | 68 | 80 | 80 | 68 | 77 | 84 | 64 | 65 | 71 |
| 94 | 97 | 100 | 82 | 80 | 77 | 85 | 90 | 98 | 98 | 67 | $f$ | 99 | 89 | 95 | 95 | 97 | 91 | 84 | 97 | 98 | 95 |
| 95 | 98 | 99 | 81 | 79 | 79 | 88 | 92 | 99 | 99 | 70 | 100 | $t$ | 89 | 97 | 96 | 98 | 93 | 87 | 98 | 99 | 95 |
| 92 | 94 | 94 | 74 | 72 | 77 | 88 | 91 | 95 | 95 | 73 | 95 | 96 | $m$ | 88 | 87 | 86 | 85 | 81 | 85 | 87 | 87 |
| 94 | 97 | 97 | 77 | 75 | 86 | 93 | 97 | 100 | 100 | 77 | 98 | 99 | 96 | $h_w$ | 100 | 97 | 98 | 96 | 94 | 95 | 96 |
| 94 | 96 | 97 | 76 | 73 | 87 | 94 | 97 | 100 | 100 | 79 | 98 | 99 | 95 | 100 | $R$ | 96 | 98 | 96 | 93 | 94 | 96 |
| 93 | 95 | 98 | 89 | 87 | 79 | 85 | 91 | 97 | 98 | 65 | 98 | 98 | 93 | 97 | 97 | $\hbar$ | 94 | 88 | 99 | 99 | 95 |
| 92 | 91 | 94 | 79 | 76 | 89 | 91 | 95 | 97 | 97 | 70 | 95 | 96 | 94 | 97 | 97 | 96 | $\pi$ | 97 | 91 | 91 | 93 |
| 92 | 93 | 92 | 69 | 66 | 91 | 98 | 100 | 98 | 98 | 84 | 93 | 95 | 93 | 98 | 98 | 93 | 96 | $e$ | 82 | 84 | 88 |
| 92 | 94 | 98 | 91 | 88 | 75 | 82 | 88 | 96 | 96 | 62 | 98 | 98 | 93 | 96 | 96 | 100 | 95 | 91 | $s$ | 99 | 94 |
| 94 | 96 | 99 | 88 | 85 | 76 | 83 | 89 | 97 | 97 | 63 | 99 | 99 | 94 | 97 | 97 | 99 | 95 | 92 | 100 | $h_T$ | 95 |
| 90 | 91 | 96 | 88 | 86 | 82 | 84 | 90 | 96 | 96 | 62 | 96 | 96 | 93 | 96 | 95 | 98 | 98 | 91 | 98 | 97 | $x$ |

**Fig. 1** Citation records for 6 physicists with about the same number of publications. Connecting lines are plotted as guides for the eye. Big symbols indicate (from left to right) the $w$ index, the $h_2$ index, the $h$ index, the number $n_1$ of publications which are cited at least once and the total number $n$ of publications. The intersection of the connecting lines with the steep straight (broken) line given by $c(r) = 10r$ yields the $w$ index, with the (short-dashed) parabola given by $c(r) = r^2$ determines the $h_2$ index, and with the more shallow straight (solid) line given by $c(r) = r$ establishes the $h$ index. Due to the restriction of the indices to integer values the respective large symbols do not lie on the interpolating connecting lines, but rather below (often significantly) on the straight lines or the parabola, if only the inequalities and not the equality in Eqs. (1-3) are fulfilled. The intersections of the flat straight solid line with the connecting lines yield the interpolated values of the index $\tilde{h}$ according to Eq. (7). Four data points, namely $c^J(1)$, $c^P(1)$, $c^P(2)$, $c^P(3)$, are not shown, because they lie outside the displayed range.

**Fig. 2** Average citation counts for the same datasets as in Fig. 1. Connecting lines are plotted as guides for the eye. Big symbols indicate (on the curves from left to right) the highest number of citations $c_1$, the average number of citations in the elite set $\bar{c}(n_\pi)$, the $A$ index, the $\tilde{g}$ index, and the average number $\bar{c}(n)$ of citations to all papers. The large symbols which do not lie on any line, but rather left of and above the straight line as well as left of and below the data curves reflect the $m$ index plotted at rank $h$. The intersections of the connecting lines with the straight line given by $\bar{c}(r) = r$ yield the interpolated values of the index $\tilde{g}$ according to Eq. (14). Further large symbols on the straight line show (from left to right) the $f$ index, the $t$ index, and the $g$ index. (Some of these values coincide so that not all symbols can be distinguished.) For the $g$ index these symbols are usually slightly below the respective values of the functions $\bar{c}(g)$ because only integer values are possible for the $g$ index, so that only the inequalities and not the equality in Eq. (13) is fulfilled.

**Fig. 3** Square root of the summed citation counts for the same datasets as in Figs. 1 and 2. Connecting lines are plotted as guides for the eye. Big symbols indicate (on the curves from left to right) the $h_w$ index, the $R$ index, the $\tilde{g}$ index, and $\sqrt{2}\,\hbar$. The intersections of the connecting lines with the straight line given by $\sqrt{s(r)} = r$ yield the interpolated values of the index $\tilde{g}$ according to Eq. (26). The further large symbols on the straight line which show the $g$ index are usually below the respective values of the functions $\sqrt{s(g)}$ because only integer values are possible for the $g$ index, so that only the inequalities and not the equality in Eq. (25) is fulfilled.

Figure 1

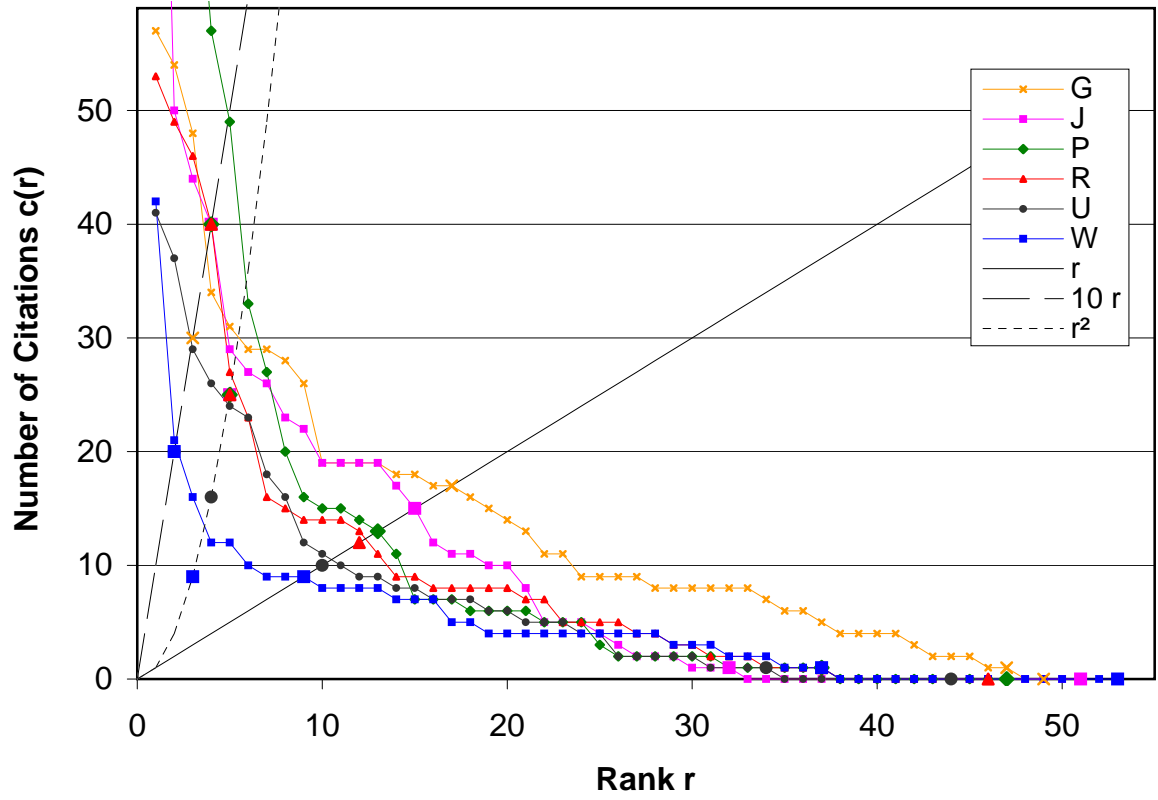

Figure 2

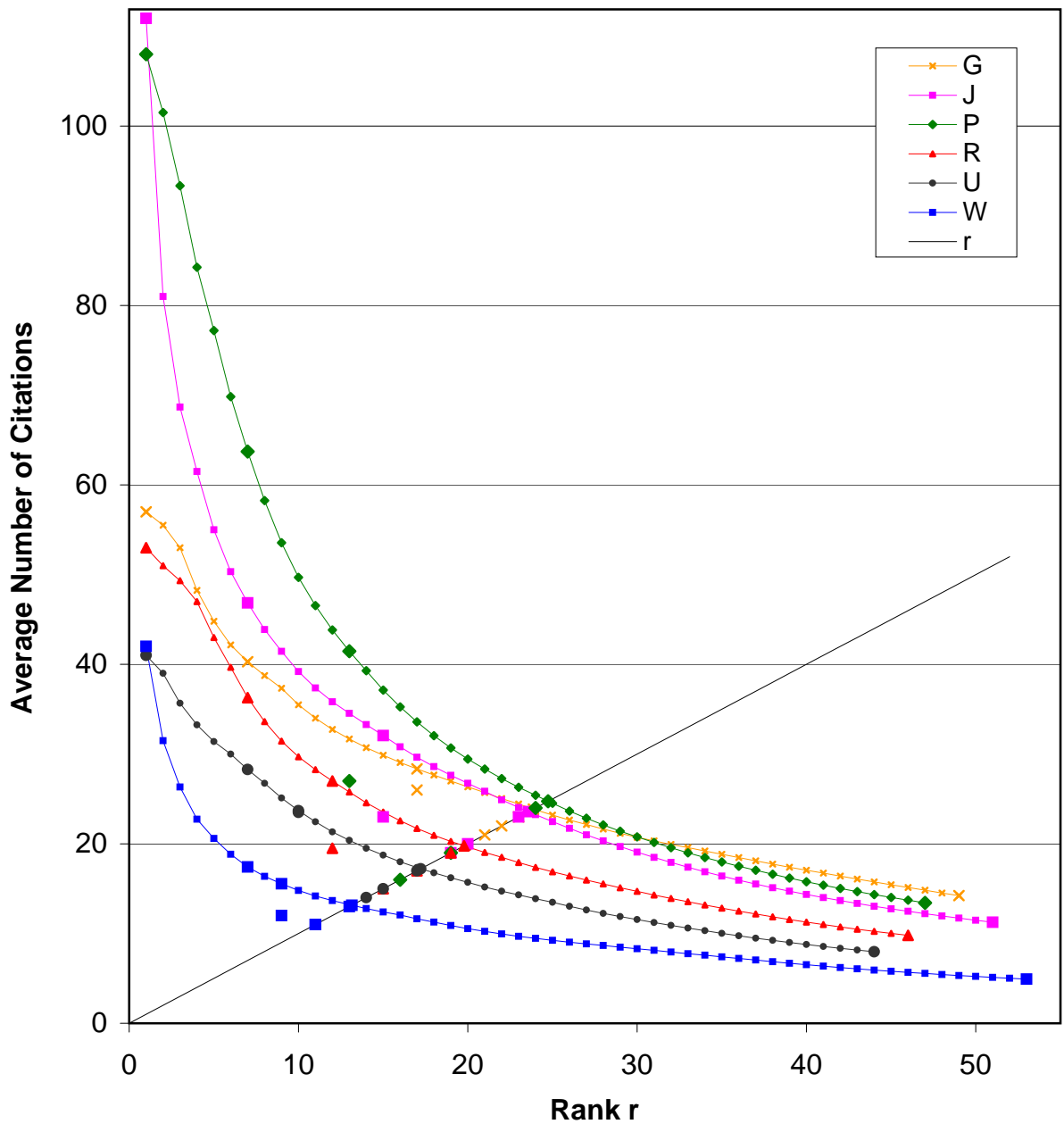

Figure 3

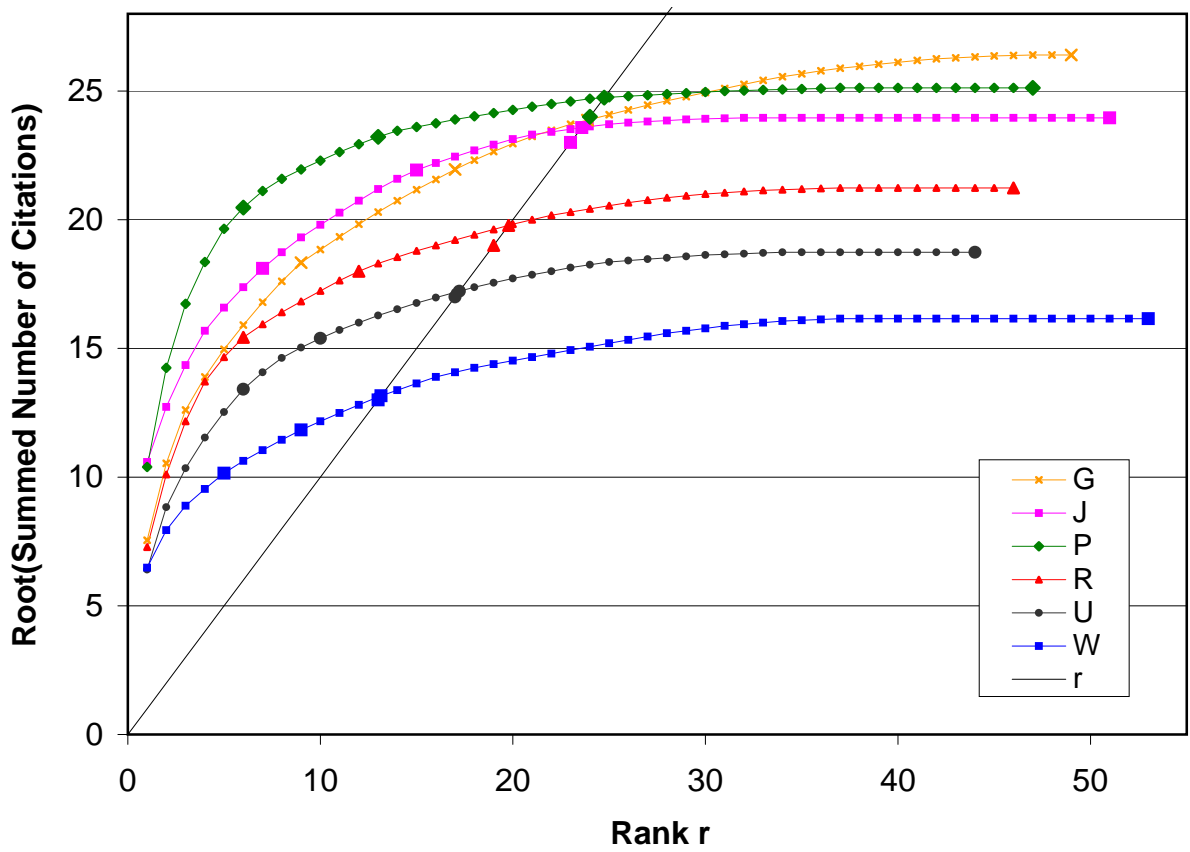